# The Role of annealed defects on conformational statistics of a self-avoiding semi-flexible polymer chain: Exact Results (I)

Pramod Kumar Mishra
Email: pkmishrabhu@gmail.com

*Department of Physics, DSB Campus, Kumaun University, Nainital (Uttarakhand) India-263002*

***Abstract*—** We study equilibrium statistics of single semi-flexible polymer chain in the presence of defects. The defects are lying along a line in the two and three dimensions and the monomers are interacting with the onsite potential of the defects. A fully directed self-avoiding walk model is used in two and three dimensions to describe thermo-dynamical behaviour of the chain in the presence of *m* defects. We have found that the number of conformations of the semi-flexible polymer chain may be controlled by means of introduction of such defects so that a particular fraction of the chain conformations may be either suppressed or populated as per our requirements for synthesizing the polymer-nano-aggregates. We discuss the role of annealed defects for its *Q* realizations and *m* defects analytically, i. e. when the defects are in the thermal equilibrium with the monomers of the semi-flexible chain.

## 1. INTRODUCTION

Recent advances in the experimental techniques made it now possible to study structural aspects of single macromolecule through devices like the optical tweezers, the atomic force microscope (AFM). The techniques based on AFM are used to obtain elastic constants and the thermo-dynamical parameters of the single macromolecule experimentally [1-6]. The conformations of the macromolecule (*DNA*, *Proteins*) play vital role in the Biological processes; and study of the conformational behaviour of the macro-molecules may have several useful technological applications, too. Some of the applications of the polymeric materials may be steric stabilization of the dispersions of the polymer solutions, functioning of the Biosensors, surface coatings using polymeric materials [7-10], and the controlled drug delivery using Biodegradable matrix (nano-polymer aggregate) of the polymer to transport drug-compounds to the actual site of living cells or the organs. We consider defects located along a line to describe behaviour of a linear homo-polymer chain aggregated in the vicinity of the defects.

There are limited reports regarding statistics of a polymer chain (either for the flexible or for the semi-flexible chain) containing defects and there are reports on the ideal polymer chain with defects, and the reports are available in the literature for the ideal polymers, see [11-14]; and references quoted therein. In other words, the equilibrium statistics is least understood for the self-avoiding polymer chain with defects. Therefore, we consider a self-avoiding homo-polymer chain to study behaviour of the chain in the presence of the *m* defects; say, $m$ $(\geq 1)$ non-interacting or uncorrelated defects lying along a line. The defects are in the thermal equilibrium with the monomers of the chain, and thus in the present report we consider case of the annealed defects, and the quenched defects case had been planned to be published elsewhere. Since, *m*-defects are located along a line in two and three dimensional space; and therefore the conformations of the chain had been enumerated using fully directed self avoiding walk model [15-21] on a square and the cubic lattices for the sake of simplicity to discuss behaviour of the semi-flexible chain in the two as well as three dimensions in the presence of *m* annealed defects. The thermodynamics of defected chain with quenched defects may be studied using method described by the author in his recent paper [22].

The paper is organized as follows: we describe fully directed self-avoiding walk model in brief for the semi-flexible polymer chain in the section two. The bending of the polymer chain may also be treated as the defects in the chain. A Boltzmann weight of bending energy for each bend in the chain may mimic semi-flexible polymer chain. The section three illustrates the methods of calculations of the thermo-dynamical parameters for the defected semi-flexible polymer chain. We summarize our findings in the section four and conclude the discussion by highlighting our findings in the section five.

## 2. THE MODEL

The fully directed self-avoiding walk model [15-21] has been used to mimic conformations of a defected semi-flexible homo-polymer chain in two and three dimensions. A conformation of seven monomers long defected polymer chain is shown schematically in the figure no.1. The chain is grafted at a point O and defects are located along a line (say, $y=0$). The defect is shown as the two closed circles connected through solid line; while the monomers of the chain is shown as the two open circles connected through a solid line. The walker is allowed to take steps only along $+x$ and $+y$ directions in the case of square lattice; while in the case of cubic lattice the walker is allowed to take steps along $+x$, $+y$ and $+z$ directions, only. Each bend of the chain is included in the model by introducing a Boltzmann weight [$k=Exp(-\varepsilon_B/k_BT)$] for the bending energy ($\varepsilon_B$) to mimic behaviour of the semi-flexible defected polymer chain [15-22].

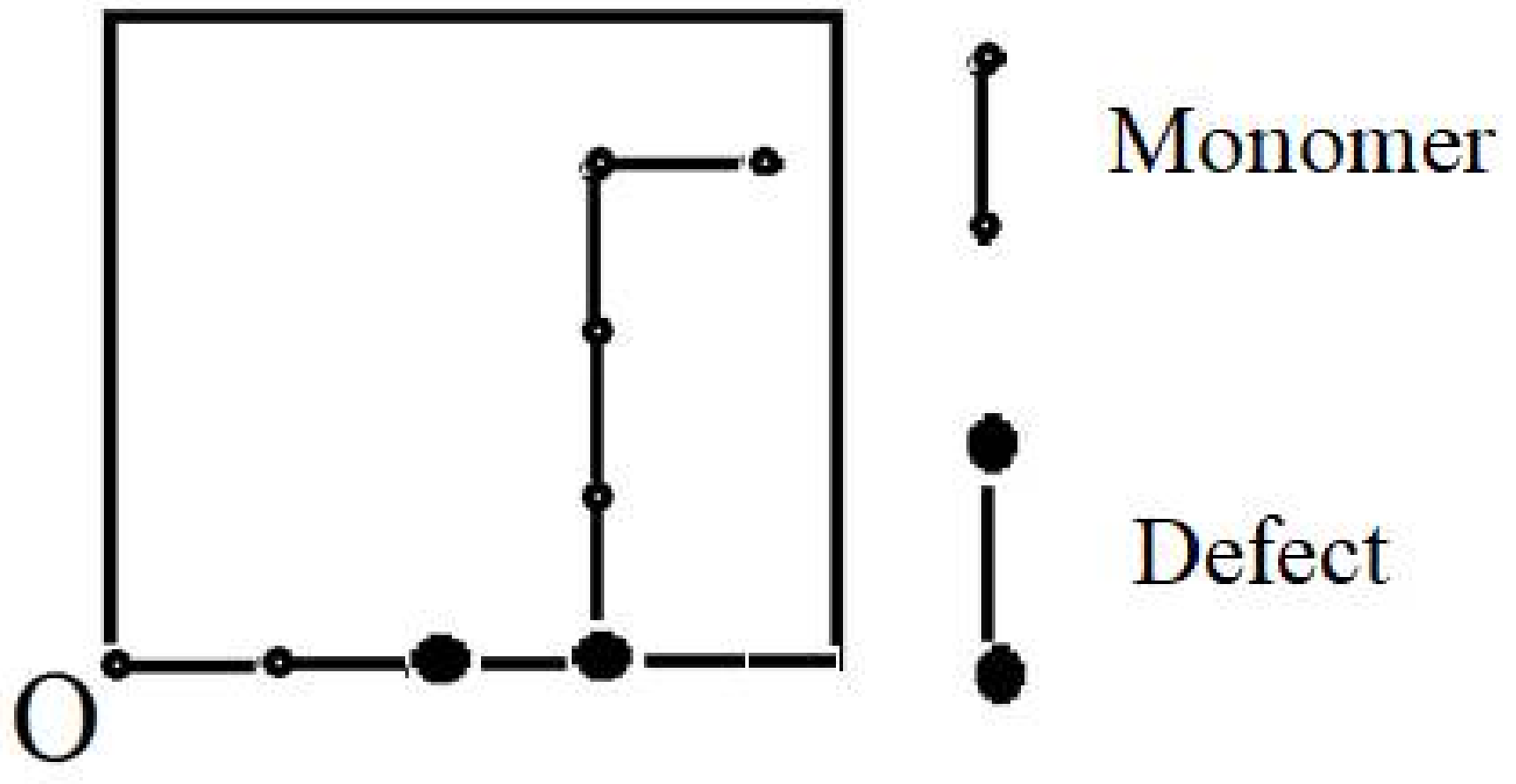

**Figure No. 1:** We have shown a conformation of seven monomers long semi-flexible polymer chain schematically containing two bends; and a defect is also shown in form of two closed circles connected through a solid line. The defect is located two monomers distant from a point O. One end of the chain is grafted at the point O.

The grand canonical partition function of the defected polymer chain is obtained using methods of recursion relations [15-22]; and the actual contributions were calculated separately for defected polymer chain and general expression of the partition function for two as well three dimensions may be written as (where in the d-dimensions, there are (d-1)*$d^{N-1}$ conformations of the $N$ monomers long flexible chain are defect free and it is due to fact that defects are lying along a line, y=0; and chain may not revisit the defect line once it moves away from the line),

$$G(x,g,k)=\sum_{P=1}^{N\to\infty}\sum_{K=0}^{\infty}\sum_{N_B=0}^{N-1}x^P g^K k^{N_B} \tag{1}$$

Where $x$ (=$u*g$) and $g$ are the step fugacity of the walk along the line of the defect and in the bulk, respectively. The Boltzmann weight corresponding to bending energy of the chain is shown using stiffness factor $k$; the term $u$ is the Boltzmann weight corresponding to onsite potential due to the defect (and the onsite potential of the defect may be attractive, repulsive or zero), $N_B$ is the number of bends in the chain; while $P$ is the position of the defect along the chosen line of the defects, and $K=N-P$, $(P\leq N)$.

## 3. THE METHOD OF CALCULATIONS

We enumerated conformations of the defected chain in the two and three dimensions using fully directed self avoiding walks on a square lattice and the cubic lattices, respectively; and the defect is lying along a line i. e. (say, $y=0$); and there are $N_B$ bends in the chain's conformations and the Boltzmann weight for the possible defected conformations is written as,

$$G(x,g,k)=\sum_{P=1}^{N\to\infty}\sum_{K=0}^{N-P}C(P,K,k)g^K x^P \tag{2}$$

Where, $d^{N-1}$ is the maximum number of conformations of $N$ monomers long flexible chain which are defected; and thus we have sum of the Boltzmann weight of the defected exact number of the conformations [$C(P,K,k)$] for $N$ (=$P+K$, here $P$ is the position of the defect along the line, and while $K$ is the number of monomers in the bulk) monomers long defected chain; where $P\leq;N$ and $P=1,2,3\ldots,N$.

$$C(P,K,k)=1+\sum_{P}^{N}[\sum_{M=1}^{K}k^M\frac{\prod_{M=1}^{K}(K-M-1)}{(M-1)!}] \tag{3}$$

$$C(P,K,k)=1+\sum_{P}^{N}\{\sum_{M=1}^{K}[C_{M-1}^{K-1}k^M]\} \tag{4}$$

However, in the case of three dimensions, the conformations of the defected chain may be counted using the following relations and correspondingly the Boltzmann weight or the canonical partition function of the chain may be obtained using standard methods of statistical Physics. It is possible to get the canonical partition function of an $N$ monomers long chain in two and three dimensions; $G(N,k)=(d-1)*[1+\sum_{P=1}^{N}C(P,K,k)]$ when the semi-flexible chain is free from the defects.

The number of defected chain conformations in three dimensions is written as:

$$C(P,K,k) = 1 + \sum_{P}^{N}\{\sum_{M=1}^{K} 2^M \frac{(K-1)(K-2)...\{K-(M-1)\}}{(M-1)!} k^M\} \quad (5)$$

$$C(P,K,k) = 1 + \sum_{P}^{N}\{\sum_{M=1}^{K} 2^M C_{M-1}^{K-1} k^M\} \quad (6)$$

Where, $M \geq 1$, $K(=N-P$, monomers are in the bulk) $\geq 1$; $P \leq N$ (and $P$ monomers are along the line of defects), and $N_B \leq (N-1)$; we may introduce single to $P$ number of defects in the chain, and since the defected polymer chain is $N$ monomers long, and index P bear information regarding position of the defect in the space and thus C(P, K, k) is the number of polymer chain conformations of N monomers containing $N_B$ bends in the C(P, K, k) conformations; {C(P,K, k)=1, when $N_B$=0}. For the convenience, the number of defected conformations is shown in the table no. 1 for the possible position of a defect in a semi-flexible chain of $N$=100 monomers.

The defects and the monomers of the chain may be in the thermal equilibrium (annealed disorder) and the *Helmholtz* free energy of the polymer chain containing annealed defects may be written using following equation; where only those conformations were dominant contributor in the partition function of the chain that contains defect(s) in its segment:

$$F_A(x,g,k) = -k_B T * Log[\frac{1}{Q} \sum_{All-walks-of-N-monomers-for-Q-realizations} C(P,N,k) \quad ] \quad (7)$$

Here, $k_B$ is the Boltzmann constant. It is to be noted that in the case of q defects in the N monomers long chain there are possible Q realizations of the defects in the chain and Q= $\frac{N(N-1)(N-2)...(N-q-1)}{q-1}$. Since each realization corresponds to the conformations of N monomers long defected chain and hence scaled free energy in the thermodynamic limit for $Q$ realizations may be written as,

$$\frac{F_A(x,g,k)}{k_B T} = -Log[g^N * f_N(k)] \quad (8)$$

The stiffness dependent term $f_N(k)$ may be obtained using following relation,

$$f_N(k) = \frac{\sum_{All-walks-of-Nmonomers-for-Q-realizations} C(P,N,k)}{\sum_{All-walks-of-Nmonomers-for-Q-realizations} C(P,N,k=1)} \quad (9)$$

The thermo-dynamical properties of the defected semi-flexible chain may be calculated using above equation nos. (7) (8) and (9) provided the length of the polymer chain may be larger in comparison to number of realizations ($Q$) of the $m$ defects distributions; and it is merely because one end of the chain is grafted at a point O. The methods described in the present report may be extended to other cases of self-avoiding polymer chain in the disordered environment and therefore it may be a useful analytical approach to discuss conformational statistics of the defected semi-flexible chain.

| N | P | $C^{2d}_{P,N,k}$ | $C^{3d}_{P,N,k}$ |
|---|---|---|---|
| 100 | 1 | $kC_0^{98} + k^2C_1^{98} + k^3C_2^{98} + k^4C_3^{98} + k^5C_4^{98} + \cdots\ldots + k^{99}C_{98}^{98}$ | $2kC_0^{98} + 2^2k^2C_1^{98} + 2^3k^3C_2^{98} + 2^4k^4C_3^{98} + 2^5k^5C_4^{98} + \cdots\ldots + 2^{99}k^{99}C_{98}^{98}$ |
| | 2 | $kC_0^{97} + k^2C_1^{97} + k^3C_2^{97} + k^4C_3^{97} + k^5C_4^{97} + \cdots\ldots + k^{98}C_{97}^{97}$ | $2kC_0^{97} + 2^2k^2C_1^{97} + 2^3k^3C_2^{97} + 2^4k^4C_3^{97} + 2^5k^5C_4^{97} + \cdots\ldots + 2^{98}k^{98}C_{97}^{97}$ |
| | 3 | $kC_0^{96} + k^2C_1^{96} + k^3C_2^{96} + k^4C_3^{96} + k^5C_4^{96} + \cdots\ldots + k^{97}C_{96}^{96}$ | $2kC_0^{96} + 2^2k^2C_1^{96} + 2^3k^3C_2^{96} + 2^4k^4C_3^{96} + 2^5k^5C_4^{96} + \cdots\ldots + 2^{97}k^{97}C_{96}^{96}$ |
| | 4 | $kC_0^{95} + k^2C_1^{95} + k^3C_2^{95} + k^4C_3^{95} + k^5C_4^{95} + \cdots\ldots + k^{96}C_{95}^{95}$ | $2kC_0^{95} + 2^2k^2C_1^{95} + 2^3k^3C_2^{95} + 2^4k^4C_3^{95} + 2^5k^5C_4^{95} + \cdots\ldots + 2^{96}k^{96}C_{95}^{95}$ |
| | 5 | $kC_0^{94} + k^2C_1^{94} + k^3C_2^{94} + k^4C_3^{94} + k^5C_4^{94} + \cdots\ldots + k^{95}C_{94}^{94}$ | $2kC_0^{94} + 2^2k^2C_1^{94} + 2^3k^3C_2^{94} + 2^4k^4C_3^{94} + 2^5k^5C_4^{94} + \cdots\ldots + 2^{95}k^{95}C_{94}^{94}$ |
| | 6 | $kC_0^{93} + k^2C_1^{93} + k^3C_2^{93} + k^4C_3^{93} + k^5C_4^{93} + \cdots\ldots + k^{94}C_{93}^{93}$ | $2kC_0^{93} + 2^2k^2C_1^{93} + 2^3k^3C_2^{93} + 2^4k^4C_3^{93} + 2^5k^5C_4^{93} + \cdots\ldots + 2^{94}k^{94}C_{93}^{93}$ |
| | 7 | $kC_0^{92} + k^2C_1^{92} + k^3C_2^{92} + k^4C_3^{92} + k^5C_4^{92} + \cdots\ldots + k^{93}C_{92}^{92}$ | $2kC_0^{92} + 2^2k^2C_1^{92} + 2^3k^3C_2^{92} + 2^4k^4C_3^{92} + 2^5k^5C_4^{92} + \cdots\ldots + 2^{93}k^{93}C_{92}^{92}$ |
| | 8 | $kC_0^{91} + k^2C_1^{91} + k^3C_2^{91} + k^4C_3^{91} + k^5C_4^{91} + \cdots\ldots + k^{92}C_{91}^{91}$ | $2kC_0^{91} + 2^2k^2C_1^{91} + 2^3k^3C_2^{91} + 2^4k^4C_3^{91} + 2^5k^5C_4^{91} + \cdots\ldots + 2^{92}k^{92}C_{91}^{91}$ |
| | 9 | $kC_0^{90} + k^2C_1^{90} + k^3C_2^{90} + k^4C_3^{90} + k^5C_4^{90} + \cdots\ldots + k^{91}C_{90}^{90}$ | $2kC_0^{90} + 2^2k^2C_1^{90} + 2^3k^3C_2^{90} + 2^4k^4C_3^{90} + 2^5k^5C_4^{90} + \cdots\ldots + 2^{91}k^{91}C_{90}^{90}$ |
| | 10 | $kC_0^{89} + k^2C_1^{89} + k^3C_2^{89} + k^4C_3^{89} + k^5C_4^{89} + \cdots\ldots + k^{90}C_{89}^{89}$ | $2kC_0^{89} + 2^2k^2C_1^{89} + 2^3k^3C_2^{89} + 2^4k^4C_3^{89} + 2^5k^5C_4^{89} + \cdots\ldots + 2^{90}k^{90}C_{89}^{89}$ |
| | . . . 20 | . . . $kC_0^{79} + k^2C_1^{79} + k^3C_2^{79} + k^4C_3^{79} + k^5C_4^{79} + \cdots\ldots + k^{80}C_{79}^{79}$ | . . $2kC_0^{79} + 2^2k^2C_1^{79} + 2^3k^3C_2^{79} + 2^4k^4C_3^{79} + 2^5k^5C_4^{79} + \cdots\ldots + 2^{80}k^{80}C_{79}^{79}$ |
| | . . . 30 | . . . $kC_0^{69} + k^2C_1^{69} + k^3C_2^{69} + k^4C_3^{69} + k^5C_4^{69} + \cdots\ldots + k^{70}C_{69}^{69}$ | . . $2kC_0^{69} + 2^2k^2C_1^{69} + 2^3k^3C_2^{69} + 2^4k^4C_3^{69} + 2^5k^5C_4^{69} + \cdots\ldots + 2^{70}k^{70}C_{69}^{69}$ |
| | . . . 100 | . . . $k^0$ | . . $k^0$ |
| Total conformations | | $2^{99}$ | $3^{99}$ |

It is also to be noted that once the free energy of the defected chain is known other thermo-dynamical parameter of the defected semi-flexible chain may follow easily from the following equations:

$$< N >= -\frac{\partial Log(\frac{F}{k_B T})}{\partial Log(g)} \quad (10)$$

$$< N >= NLog(g_N^{-1}) - \frac{1}{f_N(k)}\frac{df_N(k)}{dN} \quad (11)$$

$$< N_B >= -\frac{\partial Log(\frac{F}{k_B T})}{\partial Log(k)} \quad (12)$$

$$< N_B >= N\frac{d[\mu(N)}{dk} - \frac{df_N(k)}{dk} \quad (13)$$

We define persistence length ($l_P$) as the average length of the chain in between its two successive bends [18-22],

$$l_P = \frac{NLog[\mu(N)] - \frac{1}{f_N(k)}\frac{df_N(k)}{dN}}{N\frac{d[\mu(N)]}{dk} - \frac{df_N(k)}{dk}} \quad (14)$$

## 4. THE SUMMARY AND DISCUSSIONS

The equilibrium statistics of a defected semi-flexible chain is described using lattice model of the fully directed self-avoiding semi-flexible polymer chain when the chain is interacting with the defects. The defects are lying along a line and the number realizations of the defects (i. e. *Q* realizations) were considered for the case of *m* annealed defects in the space (i. e. in the two and three dimensions). A lattice model of fully directed self-avoiding walk is used to calculate exact number of the defected conformations analytically. We have considered very simple method to estimate the actual number of defected conformations of the semi-flexible chain of given length (N), where $N_B$ bends are seen in the conformations of the chain. There are *m* number of defects may be located along the length of chain and actual number of conformations may be obtained using method outlined in the present report and hence accordingly one may obtain free energy and other required thermo-dynamical parameters of the defected semi-flexible polymer chain for the disordered case.

The method of calculation of the free energy of the chain containing defects along its length may be used for several realizations (*Q*) of the uncorrelated *m*-defects lying on the semi-flexible chain; and therefore present report may be useful to find statistics of the defected semi-flexible chain in the form of nano-polymer-aggregates. However, the effect of such defects are marginal due to mathematical simplicity introduced in the proposed fully directed walk model system of the chain and also because of the consideration of the defects lying along a line.

## 5. CONCLUSIONS

We derive equilibrium statistics of a semi-flexible polymer chain in the presence of annealed defects. The defects are *m* in numbers and lying along a line in two and three dimensions. We have considered the fully directed self-avoiding walk model for the linear homo-polymer chain to describe the thermodynamics of the defected semi-flexible chain analytically. The method described in the present report may be useful to predict thermo-dynamical behaviour of a finite and an infinite semi-flexible defected polymer chain for *m* defects; and also for the *Q* realizations where the chain may appear in the form of nano-polymer-aggregates.

**Acknowledgements:** Author would like to thank his mentor Professor Yashwant Singh, Institute of Science, Banaras Hindu University, Varanasi (U. P.) India, for the useful discussions & training.
**About the Author: Dr. Pramod Kumar Mishra (**also know as **P. K. Mishra);**
Publons-id: http://publons.com/a/1336175/
Orcid id: http://orcid.org/0000-0003-4640-2374